\documentclass[preprint,showpacs,preprintnumbers,superscriptaddress,amsmath,amssymb]{revtex4}

\usepackage{graphicx}
\usepackage{dcolumn}
\usepackage{bm}

\begin{document}

\title{Evidence of local structural inhomogeneity in FeSe$_{1-x}$Te$_{x}$ from 
extended x-ray absorption fine structure}
\author{B. Joseph}
\author{A. Iadecola}
\affiliation{Dipartimento di Fisica, Universit\`{a} di Roma ``La 
Sapienza", P. le Aldo Moro 2, 00185 Roma, Italy}
\author{A. Puri}
\affiliation{Dipartimento di Fisica, Universit\`{a} di Roma ``La 
Sapienza", P. le Aldo Moro 2, 00185 Roma, Italy}
\affiliation{INFN - Laboratori Nazionali di Frascati, Via E. Fermi 40, 
00044 Frascati, Rome, Italy}
\author{L. Simonelli}
\affiliation{European Synchrotron Radiation Facility, 6 RUE Jules 
Horowitz BP 220 38043 Grenoble Cedex 9 France}
\author{Y. Mizuguchi}
\author{Y. Takano}
\affiliation{National Institute for Materials Science, 1-2-1 Sengen, 
Tsukuba, 305-0047, Japan\\
JST-TRIP, 1-2-1 Sengen, Tsukuba 305-0047, Japan}
\author{N.L. Saini}
\email[Corresponding author: ]{Naurang.Saini@roma1.infn.it}
\affiliation{Dipartimento di Fisica, Universit\`{a} di Roma ``La 
Sapienza", P. le Aldo Moro 2, 00185 Roma, Italy}

\begin{abstract}
Local structure of FeSe$_{1-x}$Te$_{x}$ has been studied by extended
X-ray absorption fine structure (EXAFS) measurements as a function of
temperature.  Combination of Se and Fe $K$-edge EXAFS has permitted to
quantify the local interatomic distances and their mean-square
relative displacements.  The Fe-Se and Fe-Te bond lengths in the
ternary system are found to be very different from the average
crystallographic Fe-Se/Te distance, and almost identical to the Fe-Se
and Fe-Te distances for the binary FeSe and FeTe systems, indicating
distinct site occupation by the Se and Te atoms.  The results provide
a clear evidence of local inhomogeneities and coexisting electronic
components in the FeSe$_{1-x}$Te$_{x}$, characterized by different
local structural configurations, with direct implication on the
fundamental electronic structure of these superconductors.
\end{abstract}


\maketitle

Among others, an important development in the field of Fe-based
superconductors \cite{rev_hosono} has been the discovery of
superconductivity in the FeSe(Te) chalcogenides with the maximum
T$_{c}$ $\sim$15 K \cite{dis_FeSe,dis_FeSeTe1,dis_FeSeTe2}.  Indeed,
the PbO-type tetragonal phase of FeSe(Te) has an Fe based planar
sublattice, similar to the layered FeAs-based pnictide structures with
stacking of edge-sharing Fe(Se,Te)$_{4}$-tetrahedra layer by layer.
Apart from the similar structural topology of active layers, the
chalcogenides show structural transition from tetragonal to
orthorhombic phase \cite{PT_FeSe_McQueenPRBPRL,PT_FeTe_BaoPRL,
PT_Horigane,PT_FeSe} analogous to that observed in the
FeAs-based pnictides \cite{rev_hosono,Fratini}.  Also, the interplay
between the superconductivity and itinerant magnetism, observed in
FeAs-based pnictides
\cite{rev_hosono,GokoUemura_PRB,DaGorkov1,DaGorkov2}, is well extended
to the chalcogenides
\cite{Bozin_PRB,Pres_Medvedev,Mag_SC_FeSeTe_Khasanov,mag_sc_fesete2}.
Therefore, the FeSe(Te) system provides a unique opportunity to study
the interplay of the structure, magnetism and superconductivity in the
Fe-based families because of the relative chemical simplicity with
added advantage of the absence of any spacer layers, that may affect
the electronic and structural properties within the active Fe-Fe
layers \cite{Fratini}.

One of the particularly interesting aspects of chalcogenides is the
strong relationship between the superconducting state and the defect
chemistry \cite{PT_FeSe_McQueenPRBPRL,PT_FeSe}.  In addition, the
superconducting state can be manipulated by the pressure (either
chemical \cite{Chem_press_takano,dis_FeSeTe1,dis_FeSeTe2} or
hydrostatic
\cite{Pres_Medvedev,Pres_mizuguchi,Pres_margadonna,Pres_horigane}).
Over the above, fundamental electronic and magnetic properties are
found to show extreme sensitivity to the atomic positions
\cite{ht_theory2,LDA_FeSeTe} as in the pnictides
\cite{LDA_FeSeTe,VildoPRB,tight_bind}.  Therefore, information
obtained by diffraction techniques on the average ordered structure of
FeSe$_{1-x}$Te$_{x}$ is not enough to explain the basic electronic
properties, and the knowledge of the local structure gets prime
importance.

Here, we have used extended X-ray absorption fine structure (EXAFS), a
fast and site-specific experimental tool \cite{Konings} to probe the
local structure of FeSe$_{1-x}$Te$_{x}$ system.  For the purpose, we
have combined Se and Fe $K$-edge EXAFS as a function of temperature.
The results show that the local environment around the Fe is not
homogeneous in the ternary FeSe$_{1-x}$Te$_{x}$ system, with the Fe-Te
and Fe-Se bonds being very different.  While the local structure of
the FeSe$_{1-x}$ system is consistent with the crystallographic
structure, the Se and Te atoms do not occupy the same atomic site in
the FeSe$_{1-x}$Te$_{x}$, breaking the average crystal symmetry.

Temperature dependent X-ray absorption measurements were performed in
transmission mode on powder samples of FeSe$_{0.88}$ and
FeSe$_{0.5}$Te$_{0.5}$ (Ref. \cite{Chem_press_takano}) at the BM29 beamline
of the European Synchrotron Radiation Facility (ESRF), Grenoble, using
a double crystal Si(311) monochromator.  A continuous flow He cryostat
was used to cool the samples with a temperature control within an
accuracy of $\pm$1 K. Standard procedure was used to extract the EXAFS
from the absorption spectrum \cite{Konings}.

Figure 1 shows representatives of Fourier transform (FT) magnitudes of
the Se $K$-edge (k-range 3-18\AA$^{-1}$) and Fe $K$-edge (k-range
3-17\AA$^{-1}$) EXAFS oscillations, measured on FeSe$_{0.88}$ and
FeSe$_{0.5}$Te$_{0.5}$ samples, providing partial atomic distribution
around the Se and Fe, respectively.  The structure of FeSe$_{0.88}$ and
FeSe$_{0.5}$Te$_{0.5}$ has tetragonal symmetry at room temperature.
For the earlier, a structural transition to an orthorhombic phase
appears below $\sim$100 K
\cite{PT_FeSe_McQueenPRBPRL,PT_Horigane,PT_FeSe}.  For the Se site
(probed by Se $K$-edge), there are four Fe near neighbours at a distance
$\sim$2.4 \AA$ $ (main peak at $\sim$2 \AA$ $).  The next nearest
neighbours of Se are 8 Se(Te) and 4 Fe atoms.  Contributions of these
distant shells appear mixed, giving a multiple structured peak at
$\sim$3.0-4.5\AA$ $ (upper panel).  On the other hand, for the Fe site
(probed by Fe $K$-edge) the near neighbours are 4 Se(Te) at a distance
$\sim$2.4 \AA$ $ and 4 Fe atoms at a distance $\sim$ 2.6 \AA$ $, giving
a two peak structure at $\sim$1.5-3.0\AA$ $ (Fig.  1, lower panel),
with peaks at longer distances corresponding to longer bond-lengths.
Differences in the local structure of two samples can be well
appreciated in both Se $K$ and Fe $K$-edge data with the FTs appearing
very different for the two samples.

\begin{figure}
\input{epsf}
\epsfxsize 8.5cm
\centerline{\epsfbox{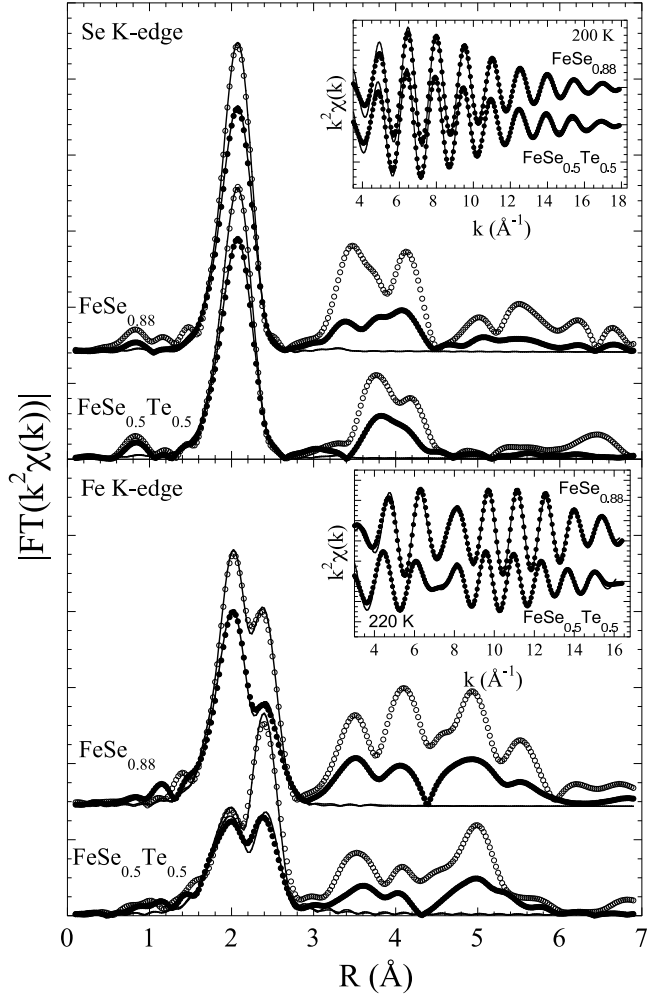}}
\caption{\label{fig:epsart} 
FT magnitudes of the Se $K$-edge (upper) and Fe
$K$-edge (lower) EXAFS oscillations at two representative temperatures,
measured (symbols) on FeSe$_{0.88}$ (35 K and 200 K) and
FeSe$_{0.5}$Te$_{0.5}$ (35 K and 220 K).  The model fits are shown as
solid lines.  The FTs are not corrected for the phase shifts, and
represent raw experimental data.  The peaks amplitudes are lower at
higher temperature (filled symbols) due to increased MSRD. The insets
show representative filtered EXAFS (symbols) with $k$-space model fits
(solid line).}
\end{figure}

\begin{figure}
\input{epsf}
\epsfxsize 8.5cm
\centerline{\epsfbox{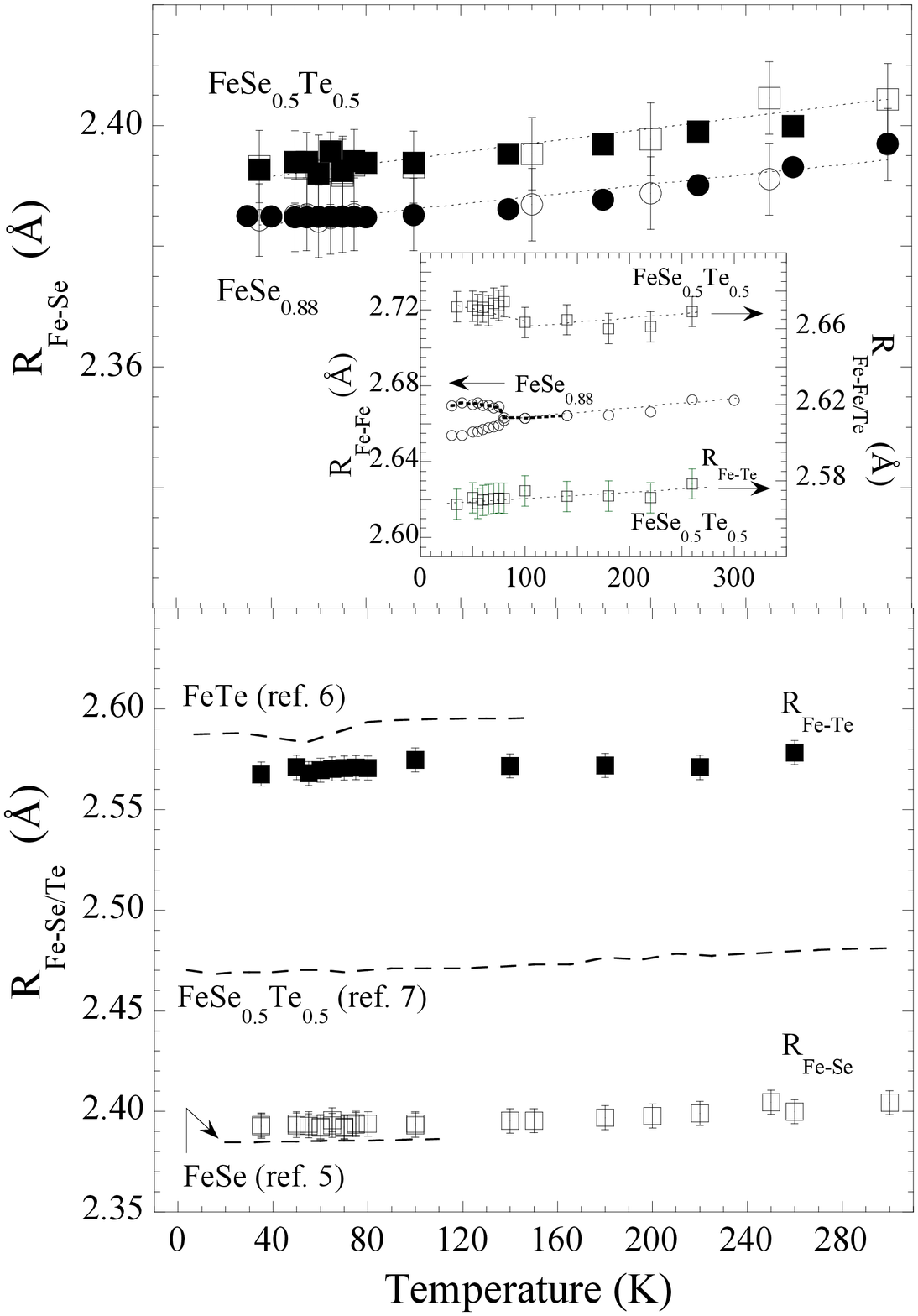}}
\caption{\label{fig:epsart}Temperature dependence of the local Fe-Se
 bond length (upper panel), determined by Se $K$-edge (open symbols) and
 Fe $K$-edge (filled symbols) for the FeSe$_{0.88}$ (circles) and
 FeSe$_{0.5}$Te$_{0.5}$ (squares).  The inset shows Fe-Fe and Fe-Te
 distances for the two samples.  The local Fe-Se (open squares) and
 Fe-Te (filled squares) distances for the FeSe$_{0.5}$Te$_{0.5}$ are 
compared with the average diffraction distances (dashed lines) in the 
lower panel.  The average Fe-Te distance in the FeTe system and Fe-Se 
distance in the FeSe system, are also included.  The uncertainties 
represent average errors, estimated by the standard EXAFS method 
using correlation maps
 (Ref. \cite{Konings}).}
\end{figure}

We will start our discussion on the Se $K$-edge since first shell EXAFS
due to Se-Fe bonds is well separated from the distant shells unlike
the Fe $K$-edge EXAFS in which the contribution of the Fe neighbours
(Se/Te and Fe) appear mixed.  From the FT itself one can appreciate
that the Se-Fe bondlengths in the sample with and without Te are
quite similar (see, e.g., upper panel of Fig.  1).  The Se-Fe EXAFS
has been analyzed using a model with a single distance.  On the other
hand, following the diffraction results
\cite{PT_FeSe_McQueenPRBPRL,PT_FeTe_BaoPRL,
PT_Horigane,PT_FeSe,Pres_Medvedev,Pres_mizuguchi,Pres_margadonna,Pres_horigane},
a three shells model was used for the Fe $K$-edge EXAFS (due to
Fe-Se(Te) bonds and Fe-Fe bonds).  The three shells for the
FeSe$_{0.88}$ are four Se at $\sim$2.4 $\AA$, and 4 Fe (2 each at
$\sim$2.6 \AA$ $ and $\sim$2.7 \AA$ $), and the same for the
FeSe$_{0.5}$Te$_{0.5}$ are two Se and two Te atoms at $\sim$2.48 $\AA$
and four Fe atoms at $\sim$2.52 \AA$ $.  The number of independent data
points: N$_{ind}\sim$ (2$\Delta k\Delta R$)/$\pi$, where $\Delta k$ and
$\Delta R$ are respectively the ranges in k and R space over which the
data are analyzed \cite{Konings}, were $\sim$19 ($\Delta k$=15
\AA$^{-1}$ and $\Delta R$=2 \AA $ $) and $\sim$17 ($\Delta k$=13
\AA$^{-1}$ and $\Delta R$=2 \AA $ $) respectively for the Se $K$-edge
(single shell fit) and the Fe $K$-edge (three shells fit) data.
Except the radial distances R$_{i}$ and related mean square relative
displacements (MSRD), determined by the correlated Debye-Waller factor
($\sigma^{2}$), all other parameters were kept constant in the
conventional least squares modelling, using the phase and amplitude
factors calculated by Feff \cite{Feff} and exploiting our experience on
the studies of similar systems \cite{loc_str_G4}.  The representative
model model fits to the Se and Fe $K$-edge EXAFS in the real and $k$-space
are also included in the Fig.  1.

Figure 2 shows temperature dependence of the Fe-Se distances
determined from the Se and Fe $K$-edge EXAFS analysis.  The Fe-Se
distance in FeSe$_{0.5}$Te$_{0.5}$ is slightly longer, however, the
temperature dependence appears similar for the two samples. Other
interatomic distances (Fig.  2 inset) are also shown with Fe-Fe
distance for the FeSe$_{0.88}$, revealing a small splitting across the
structural phase transition at $\sim$80 K. It was possible to model
the data with a single Fe-Fe distance below the structural phase
transition, however, inclusion of the two distances, consistent with
the diffraction data \cite{PT_FeSe_McQueenPRBPRL,Pres_margadonna},
improved the fit index by about 30\%.  On the other hand, Fe-Te
distance for the FeSe$_{0.5}$Te$_{0.5}$ appears only slightly shorter
than Fe-Te distance for the binary FeTe \cite{PT_FeTe_BaoPRL}.

It is interesting to note that, while the local Fe-Se distance for the
FeSe$_{0.88}$ is consistent with the average diffraction distance
\cite{PT_Horigane,Pres_margadonna}, the one for the
FeSe$_{0.5}$Te$_{0.5}$ is substantially shorter than the average
Fe-Se/Te distance (Fig.  2, lower panel).  Indeed, the local Fe-Se
distance for the FeSe$_{0.5}$Te$_{0.5}$ is quite similar to the Fe-Se
distance in the FeSe$_{0.88}$.  Also, the local Fe-Te distance in the
FeSe$_{0.5}$Te$_{0.5}$ is only slightly shorter than the average Fe-Te
distance for the binary FeTe system
\cite{PT_FeSe_McQueenPRBPRL,PT_FeTe_BaoPRL,PT_Horigane}.  The fact
that the local Fe-Se (Fe-Te) distance in the FeSe$_{0.5}$Te$_{0.5}$ is
almost equal or only slightly longer (shorter) than the Fe-Se (Fe-Te)
distance for the FeSe$_{0.88}$ (FeTe) system and much shorter (longer)
than the average Fe-Se/Te crystallographic distance, implies that
there should be a distribution of the Fe-chalcogen distances at the
local scale with Se and Te atoms sitting at different distances from
the Fe atoms.  This observation not only underlines diverging local
structure from the average one for the Te substituted sample, but also
constructs a clear evidence for an inhomogeneous distribution of the
Fe-Se and Fe-Te bonds and different active electronic components in
the chalcogenides.

\begin{figure}
\input{epsf}
\epsfxsize 8.5cm
\centerline{\epsfbox{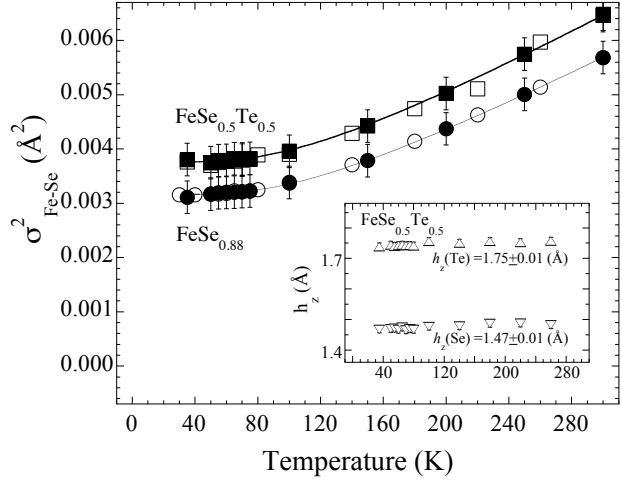}}
\caption{\label{fig:epsart} Temperature dependence of the Fe-Se MSRD
for the FeSe$_{0.88}$ (circles) and FeSe$_{0.5}$Te$_{0.5}$ (squares),
well described by the correlated Einstein model with the Einstein
temperature $\theta_{E}$=300$\pm$20 K (solid lines). The inset shows 
local Fe-chalcogen heights (h$_{z}$) in the FeSe$_{0.5}$Te$_{0.5}$ 
(triangles).}
\end{figure}

Figure 3 shows the MSRD ($\sigma^{2}$) of the Fe-Se pair, describing
distance-distance correlation function (correlated Debye Waller
factors).  The EXAFS Debye Waller factors (distance broadening) are
different from those measured by diffraction (mean-square
displacements, i.e. $\sigma_{Fe}^{2}$ and $\sigma_{Se}^{2}$).
The MSRD is sum of temperature independent ($\sigma_{0}^{2}$) and
temperature dependent terms, i.e.
$\sigma_{Fe-Se}^{2}$=$\sigma_{0}^{2}$+$\sigma_{Fe-Se}^{2}$ (T).  The
MSRD of the Fe-Se pair are well described by the correlated Einstein
model \cite{Konings,loc_str_G4} with the Einstein temperature
$\theta_{E}$ = 300$\pm$20 K, similar for the two samples within the
uncertainties, suggesting similar force constant for the Fe-Se bonds
in the two samples.  The MSRD for the FeSe$_{0.5}$Te$_{0.5}$ appears
to have slightly higher static component indicating higher disorder,
consistent with local structural inhomogeneities \cite{loc_str_G4}.
It is worth mentioning that, within the experimental uncertainties,
consideration of higher order cumulants in the analysis hardly had any
influence on the observed MSRD, consistent with negligible deviation
from the Gaussian distributions.

From the measured Fe-Se/Te and Fe-Fe bondlengths, we can determine the
local chalcogen height from the Fe-plane (h$_{z}$).  Since the two
chalcogen atoms (Se and Te) occupy two distinct sites in the
FeSe$_{0.5}$Te$_{0.5}$, there are two corresponding h$_{z}$, as shown
in Fig.  3 (inset).  The coexisting Fe-chalcogen heights in the
ternary system, showing hardly any temperature dependence, are
1.47$\pm$0.01 \AA { } and 1.75$\pm$0.01 \AA { } respectively for the
Se and Te atoms in the crystallographically homogeneous system.  The
local near neighbour distances and the Fe-chalcogen heights at a
representative temperature (T= 100 K) are shown in table 1 for a
quantitative comparison with the diffraction data.  Since the Fermi
surface topology strongly depends on the chalcogen height
\cite{ht_theory2,LDA_FeSeTe}, the results constructs a direct evidence
of local electronic inhomogeneity in the ternary FeSe$_{1-x}$Te$_{x}$
system, characterized by distribution of Fe-Se/Te bonds (and hence
Fe-chalcogen heights), consistent with the indications of low symmetry
structure in the Te substituted FeSe by electron
\cite{PT_FeSe_McQueenPRBPRL} and x-ray diffraction
\cite{lower_sym_FeSeTe}.

\begin{table*}
\caption{\label{tab:table1}Comparison between local and average near
neighbour distances (and chalcogen heights) in the FeSe$_{0.88}$ and
FeSe$_{0.5}$Te$_{0.5}$.}
\begin{tabular}{cccccccc}\hline
&&Local structure &&&Average structure \\
&&(EXAFS)&&&(Diffraction)\\
\hline
&R$_{Fe-Se}$(\AA)&R$_{Fe-Te}$(\AA)&h$_{z}$(\AA)&&R$_{Fe-Se/Te}$(\AA)&h$_{z}$(\AA)\\
\hline
FeSe$_{0.88}$&2.38&-&1.47&&2.387&1.46\\
FeSe$_{0.5}$Te$_{0.5}$&2.39&2.57&1.47(1.75)&&2.471&1.60\\
\hline
\end{tabular}
\end{table*}

Let us discuss possible implications of the present study on the
fundamental electronic structure of the title system.  Low energy
electronic states in the Fe-based superconductors are derived by the
five Fe $d$ orbitals with their relative positions modulated by the
anion (pnictogen or chalcogen) height
\cite{ht_theory2,LDA_FeSeTe,VildoPRB,tight_bind}.  The anion height
controls the Fermi surface topology (which has strong $k_{z}$
dispersion), through changing degeneracy between different bands (in
particular between the $d_{x^{2}-y^{2}}$ and $d_{xz}/d_{yz}$), with
a direct implication on the magnetic structure and superconductivity
of the Fe-based materials \cite{DFT_FeSeTe,KurokiPRL}.  For example,
it has been clearly shown \cite{ht_theory2} that a ($\pi$,$\pi$) like
single stripe magnetic order (similar to the Fe-pnictides) is more
favorable in the FeSe$_{1-x}$Te$_{x}$, with small Fe-chalcogen height,
unlike a ($\pi$,0) type double stripe pattern for the FeTe with
relatively high Fe-chalcogen height.  Therefore, the Fe-chalcogen
height seems to control the low energy states, correlation effects
(Coloumb screening through hybridizations) and magnetic order,
however, this alone is not able to describe the underlying physics.
In the light of the above, the observed local inhomogeneity has strong
implication on the physics of the chalcogenides, and in general on the
Fe-based superconductors.  Indeed, the local inhomogeneity,
characterized by coexiesting structural configurations with low and
high chalcogen heights, can easily reconcile not only the changing
magnetic order, but also the photoemission experiments on the
FeSe$_{1-x}$Te$_{x}$ systems \cite{XiaARPES,TamaiARPES}, debating on
surprisingly large effective electron mass and local correlation
effects.  In addition, the coexisting components can easily modify the
Fermi surface topology (and hence the interband scattering and the
nesting properties), putting strong constraints on the interpretaion
of Fermi surface topological effects and theoretical models describing
superconductivity in these materials.

Earlier, we have widely studied the copper oxide perovskites providing
clear evidence of inhomogeneous charge distribution characterized by
different local structural configurations \cite{loc_str_G4}.  The fact
that, FeSe$_{1-x}$Te$_{x}$ system manifest local structural
inhomogeneity while the superconducting T$_{c}$ is higher, provides
further indication that the local inhomogeneity should have important
role in the new Fe-based materials as well, consistent with recent
experiments in favour of mesoscopic phase separation
in the Fe-based superconductors
\cite{Mag_SC_FeSeTe_Khasanov,Bozin_PRB,GokoUemura_PRB,
DaGorkov1,DaGorkov2,pump_probe_SmFeAsO,LangPRL10}.

In conclusion, we have studied temperature-dependent local structure
of the FeSe$_{1-x}$Te$_{x}$ by combined analysis of Se and Fe $K$-edge
EXAFS. The Fe-Se (Fe-Te) bondlength for the FeSe$_{0.5}$Te$_{0.5}$ is
found to be much shorter (longer) than the average Fe-Se/Te,
indicating distinct site occupation by the Se and Te atoms and
inhomogeneous distribution of the Fe-Se/Te bonds (and hence
bond-angles) in a crystallographically homogeneous system.  The
results provide a clear evidence of local inhomogeneities with
coexisting electronic components, having direct implication on the low
lying electronic states and magnetic order, suggesting a possible
route to understand the physics of Fe-based chalcogenides.

\begin{acknowledgments}
The authors thank the ESRF staff for their cooperation during the
experiments, and A. Bianconi for the stimulating discussions and the
encouragement.
\end{acknowledgments}


\begin{thebibliography}{100}

\bibitem{rev_hosono} K. Ishida, Y. Nakai, and H. Hosono, 
J. Phys.  Soc.  Jpn. 78, 062001 (2009).

\bibitem{dis_FeSe} F.-C. Hsu, et al, 
Proc. Nat. Acad. Sci. U.S.A. 105, 14262 (2008). 

\bibitem{dis_FeSeTe1} M. H. Fang, H. M. Pham, B. Qian, T. J. Liu, E. K. Vehstedt, Y.
Liu, L. Spinu, and Z. Q. Mao, Phys. Rev. B 78, 224503 (2008).

\bibitem{dis_FeSeTe2} K.-W. Yeh, et al, 
EPL 84, 37002(2008).

\bibitem{PT_FeSe_McQueenPRBPRL} T. M. McQueen, et al, 
Phys. Rev. B 79, 014522 (2009); T. M.
McQueen, A. J. Williams, P. W. Stephens, J. Tao, Y. Zhu, V.
Ksenofontov, F. Casper, C. Felser, and R. J. Cava, Phys. Rev.
Lett. 103, 057002 (2009). 

\bibitem{PT_FeTe_BaoPRL} W.Bao, et al, 
Phys. Rev. Lett. 102, 247001 (2009).

\bibitem{PT_Horigane} K. Horigane, H. Hiraka, and K. Ohoyama, J. Phys. Soc. Jpn. 78,
074718 (2009).

\bibitem {PT_FeSe} E. Pomjakushina, K. Conder, V. Pomjakushin, M. Bendele, and
R. Khasanov, Phys. Rev. B 80, 024517 (2009).  

\bibitem{Fratini}M. Fratini, et al, Supercond.  Sci.  Technol.  21,
092002 (2008).

\bibitem{GokoUemura_PRB}T. Goko, et al, Phys.  Rev.  B 80, 024508 (2009).

\bibitem{DaGorkov1}J. T. Park, et al, Phys.  Rev.  Lett.  102, 117006 (2009).

\bibitem{DaGorkov2}K. Kitagawa, N. Katayama, H. Gotou, T. Yagi, K. Ohgushi, T.
Matsumoto, Y. Uwatoko, and M. Takigawa, Phys. Rev. Lett.
103, 257002 (2009).

\bibitem{Bozin_PRB} R. Hu, E. S. Bozin, J. B. Warren, and C. Petrovic, Phys. Rev. B
80, 214514 (2009).

\bibitem{Pres_Medvedev} S. Medvedev, et al, 
Nature Mater. 8, 630 (2009).

\bibitem{Mag_SC_FeSeTe_Khasanov} R. Khasanov, et al, 
Phys. Rev. B 80, 140511 (2009).

\bibitem{mag_sc_fesete2} J. Wen, G. Xu, Z. Xu, Z. W. Lin, Q. Li, W. Ratcliff, G. Gu, and
J. M. Tranquada, Phys. Rev. B 80, 104506 (2009).

\bibitem{Chem_press_takano} Y. Mizuguchi, F. Tomioka, S. Tsuda, T. Yamaguchi, and Y. Takano,
J. Phys. Soc. Jpn. 78, 074712 (2009).

\bibitem{Pres_mizuguchi} Y. Mizuguchi, F. Tomioka, S. Tsuda, T. Yamaguchi, and Y. Takano,
Appl. Phys. Lett. 93, 152505 (2008). 

\bibitem{Pres_margadonna} S. Margadonna, Y. Takabayashi, Y. Ohishi, Y. Mizuguchi, Y.
Takano, T. Kagayama, T. Nakagawa, M. Takata, and K. Prassides,
Phys. Rev. B 80, 064506 (2009); Chem.  Commun., 5607-5609 (2008). 

\bibitem{Pres_horigane} K. Horigane, N. Takeshita, C.-H. Lee, H. Hiraka, and K. Yamada,
J. Phys. Soc. Jpn. 78, 063705 (2009).

\bibitem{ht_theory2} C.-Y. Moon and H. J. Choi, Phys. Rev. Lett. 104, 057003
(2010). 

\bibitem{LDA_FeSeTe} T. Miyake, K. Nakamura, R. Arita, and M. Imada, J. Phys. Soc.
Jpn. 79, 044705 (2010). 

\bibitem{VildoPRB}V. Vildosola, L. Pourovskii, R. Arita, S. Biermann, and A.
Georges, Phys. Rev. B 78, 064518 (2008). 

\bibitem{tight_bind}M. J. Calderon, B. Valenzuela, and E. Bascones, Phys. Rev. B
80, 094531 (2009).

\bibitem{Konings}X-ray Absorption: Principles, Applications,
Techniques of EXAFS, SEXAFS, XANES, edited by R. Prins and D.
Koningsberger (Wiley, New York, 1988).

\bibitem{Feff}J. Mustre de Leon, J. J. Rehr, S. I. Zabinsky, and R. C. Albers,
Phys. Rev. B 44, 4146 (1991); J. J. Rehr and R. C. Albers, Rev.
Mod. Phys. 72, 621 (2000).

\bibitem{loc_str_G4} A. Bianconi, N. L. Saini, A. Lanzara, M. Missori, T. Rossetti, H.
Oyanagi, H. Yamaguchi, K. Oka, and T. Ito, Phys. Rev. Lett. 76,
3412 (1996); N. L. Saini, A. Lanzara, H. Oyanagi, H. Yamaguchi,
K. Oka, T. Ito, and A. Bianconi, Phys. Rev. B 55, 12759
(1997); N. L. Saini, A. Bianconi, and H. Oyanagi, J. Phys. Soc.
Jpn. 70, 2092 (2001); N. L. Saini, M. Filippi, H. Oyanagi, H.
Ihara, A. Iyo, and A. Bianconi, Phys. Rev. B 68, 104507 (2003);
A. Iadecola, S. Agrestini, M. Filippi, L. Simonelli, M. Fratini, B.
Joseph, D. Mahajan, and N. L. Saini, EPL 87, 26005 (2009).

\bibitem{lower_sym_FeSeTe} M. Tegel, C. Loehnert, and D. Johrendt, Solid State Commun.
150, 383 (2010).

\bibitem{DFT_FeSeTe} A. Subedi, L. Zhang, D. J. Singh, and M. H. Du, Phys. Rev. B
78, 134514 (2008).

\bibitem{KurokiPRL}
K. Kuroki, H. Usui, S. Onari, R. Arita, and H. Aoki, Phys. Rev.
B 79, 224511 (2009).

\bibitem{XiaARPES} Y. Xia, D. Qian, L. Wray, D. Hsieh, G. F. Chen, J. L. Luo, N. L.
Wang, and M. Z. Hasan, Phys. Rev. Lett. 103, 037002 (2009).

\bibitem{TamaiARPES} A. Tamai et al, Phys. Rev. Lett. 104, 097002 (2010).

\bibitem{pump_probe_SmFeAsO}T. Mertelj, V. V. Kabanov, C. Gadermaier, N. D. Zhigadlo, S.
Katrych, J. Karpinski, and D. Mihailovic, Phys. Rev. Lett. 102,
117002 (2009). 

\bibitem{LangPRL10}
G. Lang, H. J. Grafe, D. Paar, F. Hammerath, K. Manthey, G.
Behr, J. Werner, and B. Buchner, Phys. Rev. Lett. 104, 097001
(2010).

\end{thebibliography}

\end{document}